\documentclass[aip,twocolumn,letterpaper]{revtex4}

\usepackage{fullpage}

\usepackage{graphics}
\usepackage{epsfig}

\usepackage{xcolor}

\usepackage[margin=1.0in]{geometry}

\begin{document}
\title{Constraints on stellarator divertors from Hamiltonian mechanics}
\author{Allen H Boozer}
\affiliation{Columbia University, New York, NY  10027\\  ahb17@columbia.edu}

\begin{abstract}

The design of any large stellarator requires a plan for the removal of the particles and heat that are exhausted across the plasma edge.  This is called the divertor problem, for the particle exhaust must be diverted into pumping chambers.  Although the physics of diverted plasmas has many subtleties, the magnetic field configuration between the plasma edge and the surrounding chamber walls is the foundation upon which divertor design is based.  The properties of this magnetic configuration has both practical constraints and mathematical constraints from magnetic field lines obeying a 1~1/2 degree of freedom Hamiltonian.  Constraints from plasma physics will also be discussed; they need to be integrated with the constraints from from Hamiltonian mechanics in conceptual designs of stellarator divertors. 

\end{abstract}

\date{\today} 
\maketitle


\begin{figure}
\centerline{ \includegraphics[width=3.0in]{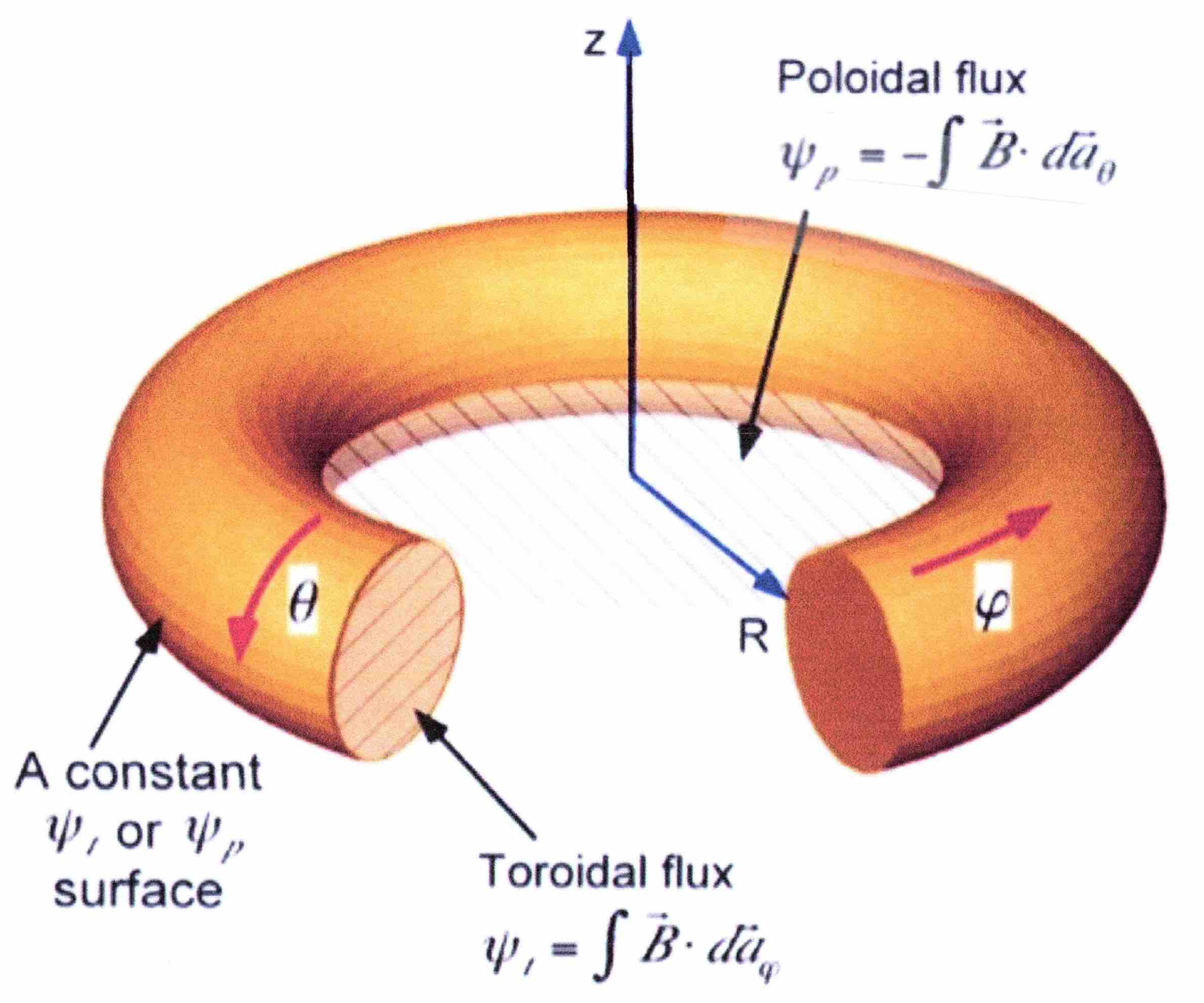} }
\caption{ The poloidal $\theta$ and the toroidal $\varphi$ angles are illustrated along with the toroidal $\psi_t$ and poloidal $\psi_p$ fluxes. }
\label{fig:Fluxes}
\end{figure}

\section{Introduction}

The plasma that crosses the outermost confining magnetic surface in a tokamak or a stellarator needs to be directed to a divertor chamber where is can be pumped away.  Two constraints are (i) the divertor chambers can cover only a small fraction of the area of the walls, and (ii) only a small fraction of the thermal energy exiting with the particles that needs to be pumped can enter the divertor.  

The second constraint requires that most of the power be radiated before the plasma enters the divertor.  The power density on the wall is equivalent to money per unit area produced by a power plant, so designs naturally make this close to the limits of materials, which is inconsistent with concentrated power striking divertors.  The required radiation can occur either in a layer that surrounds the main plasma, which is called a radiative mantle, or in a region well separated from the main plasma and close to the divertor entrance, which is called divertor detachment.

The behavior of divertors involves subtle plasma processes, which the EMC3-Eirene code \cite{EMC3-Eirene} was designed to encapsulate.  Underlying all that subtlety are constraints on the magnetic field line behavior in the region outside of the outermost magnetic surface.  General properties of the magnetic field line behavior are amenable to definite mathematical analysis since magnetic field line trajectories in tokamaks and stellarators are given exactly \cite{Boozer:1983}  by a 1~1/2 degree of freedom Hamiltonian.  This will be proven and discussed in Section \ref{B-line Ham}.  The general divertor problem resembles several well-studied problems in Hamiltonian dynamics \cite{Stochasticy:1984,Fractal-Bnd:1988,Chaos-targeting:1990,Targeting:Meiss1995}: fractal bounding surfaces for trajectories and chaos give low-energy trajectories for space probes.  Nevertheless, fundamental differences and surprises arise.

The focus of this article is on properties of the magnetic field, but a description of plasma properties that constrain practical solutions is given in Section \ref{Sec:plasma properties}.


\section{Magnetic field line Hamiltonian and chaos}

Hamiltonian mechanics and the mathematical theory of field-line chaos underlie the design of stellarator divertors.  To ensure communication, it is important to explain the central concepts in both areas.


\subsection{Magnetic field line Hamiltonian \label{B-line Ham}}

The magnetic field line Hamiltonian $\psi_p(\psi_t,\theta,\varphi)$ and its canonical variables need to be described physically, Figure \ref{fig:Fluxes}. Physically $\theta$, the canonical coordinate, is  a poloidal angle, $\varphi$, the canonical time, is a toroidal angle, $\psi_t$, the canonical momentum, is the toroidal magnetic flux enclosed by a constant-$\psi_t$ surface, and $\psi_p$, the Hamiltonian, is the poloidal magnetic flux outside of a constant-$\psi_p$ surface.  The locations of $(\psi_t,\theta,\varphi)$ points are given by a position vector $\vec{x}(\psi_t,\theta,\varphi)$, which means each Cartesian coordinate can be written as a function of $(\psi_t,\theta,\varphi)$.  The equations are
\begin{eqnarray}
2\pi \vec{B}(\vec{x}) &=& \vec{\nabla}\psi_t\times\vec{\nabla}\theta + \vec{\nabla}\varphi\times\vec{\nabla}\psi_p(\psi_t,\theta,\varphi), \label{B-rep}\\
\frac{d\psi_t}{d\varphi}&\equiv&\frac{\vec{B}\cdot \vec{\nabla}\psi_t}{\vec{B}\cdot \vec{\nabla}\varphi} = - \frac{\partial\psi_p}{\partial\theta};\\
\frac{d\theta}{d\varphi}&\equiv&\frac{\vec{B}\cdot \vec{\nabla}\theta}{\vec{B}\cdot \vec{\nabla}\varphi} = +\frac{\partial\psi_p}{\partial\psi_t}.
\end{eqnarray}

Any well behaved function, $\vec{x}(r,\theta,\varphi)$ for which a $r=$ constant surface is a torus  and $\vec{B}\cdot\vec{\nabla}\varphi \neq0$ in the region of interest can be used to define the magnetic field line Hamiltonian.  The theory of general coordinates, which is given in the appendix of \cite{Boozer:RMP} implies 
\begin{eqnarray}
\vec{\nabla}\varphi &=& \frac{ \frac{\partial \vec{x}}{\partial r} \times \frac{\partial \vec{x}}{\partial \theta} }{ \left(\frac{\partial \vec{x}}{\partial r} \times \frac{\partial \vec{x}}{\partial \theta}\right)\cdot\frac{\partial \vec{x}}{\partial \varphi}},  \hspace{0.2in}\mbox{where}\\
\mathcal{J}_r &=& \left(\frac{\partial \vec{x}}{\partial r} \times \frac{\partial \vec{x}}{\partial \theta}\right)\cdot\frac{\partial \vec{x}}{\partial \varphi}
\end{eqnarray}
is the $(r,\theta,\varphi)$ coordinate Jacobian.  An arbitrary vector in three-space can be described using three functions and the coordinate gradients, $2\pi\vec{A}(\vec{x})=\psi_t \vec{\nabla}\theta-\psi_p\vec{\nabla}\varphi + \vec{\nabla}g$.  Taking $\vec{A}$ to be the vector potential of the magnetic field $\vec{B}$, Equation (\ref{B-rep}) is obtained.  Since $2\pi\mathcal{J}_r\vec{B}\cdot\vec{\nabla}\varphi=\partial\psi_t(r,\theta,\varphi)/\partial r$, the function $\psi_t(r,\theta,\varphi)$ can found by integrating out from the $r=0$ axis with $\theta$ and $\varphi$ fixed.  The position function $\vec{x}(r,\theta,\varphi)$ can be replaced by $\vec{x}(\psi_t,\theta,\varphi)$.  The poloidal flux is found using $2\pi\mathcal{J}_r\vec{B}\cdot\vec{\nabla}\theta=\partial\psi_p(r,\theta,\varphi)/\partial r$.


\subsection{Magnetic field line chaos \label{sec:chaos} }

Magnetic field lines are said to be chaotic when neighboring lines have a separation $\vec{\delta}$ that depends exponentially on the toroidal angle $\varphi$ through which they are followed.  Neighboring means separated by an infinitesimal distance.  The exponentiation of neighboring lines can be defined line by line and at each location along the trajectories of the lines.  

Cylindrical coordinates, the $(R,\varphi,Z)$ coordinates of Figure \ref{fig:Fluxes}, give simple equations, but other coordinates could be used.  A particular magnetic field line is defined by giving its location at one point $(R_0,Z_0,\varphi_0)$.  The exponentiation in separation of the neighboring lines after the lines have been followed from $\varphi_0$ to $\varphi$ is given by
\begin{equation}
\sigma_\ell(\varphi)\equiv\sigma(R_0,\varphi_0,Z_0,\varphi),
\end{equation}
where $\sigma(R_0, \varphi_0,Z_0,\varphi)$ is the natural logarithm of the Frobenius norm of a Jacobian matrix, Equation (\ref{sigma-def}).  

The local strength of the chaos is measured by $d\sigma_\ell/d\varphi$.  The strength of chaos is conventionally measured by the Liapunov exponent, which is defined as the limit as $\varphi\rightarrow\infty$ of $\sigma_\ell(\varphi)/\varphi$, but this quantity is not useful for lines that strike the walls and does not give the local variation in the strength of the chaos.

A practical understanding of the concept of $\sigma_\ell(\varphi)$ is probably best gained by understanding how it is calculated.  The $\varphi=\varphi_0$ location of a line $(R_0,Z_0)$ defines that field line, which at $\varphi$ has the position
\begin{eqnarray}
\vec{x}(R_0,Z_0,\varphi) &=&R(R_0,Z_0,\varphi)\hat{R}(\varphi) + Z(R_0,Z_0,\varphi) \hat{Z}, \nonumber\\
&&\mbox{where} \label{Lag}\\
\frac{d\vec{x}}{d\varphi}&\equiv&\left(\frac{\partial \vec{x}}{\partial \varphi}\right)_{R_0Z_0} = \frac{\vec{B}}{\vec{B}\cdot\vec{\nabla}\varphi}.  \hspace{0.2in}\mbox{Let  }\\
 \vec{h}&\equiv&\frac{\vec{B}}{\vec{B}\cdot\vec{\nabla}\varphi},
 \end{eqnarray}
 where $\vec{B}\cdot\vec{\nabla}\varphi=RB_\varphi$ with $B_\varphi\equiv\vec{B}\cdot\hat{\varphi}$.  The function $\vec{x}(R_0,Z_0,\varphi)$ of Equation (\ref{Lag}) defines a two-dimensional Lagrangian coordinate transformation.  It gives the $(R,Z)$ position of a field line at toroidal location $\varphi$ that passed through the point $(R_0,Z_0)$ at $\varphi=\varphi_0$.

 To determine not only the field line that passes through $(R_0,Z_0)$ at $\varphi=\varphi_0$ but also all field lines in its neighborhood, two vector equations should be integrated simultaneously:
\begin{eqnarray}
\frac{d\vec{x}}{d\varphi} &=& \vec{h}(\vec{x}), \mbox{  and  } \label{x-dot} \\
\frac{d\vec{\delta}}{d\varphi} &=& \vec{\delta}\cdot\vec{\nabla}\vec{h}. \label{delta-dot}
\end{eqnarray}

Equation (\ref{x-dot}) is to be solved with the initial condition $\vec{x}(\varphi_0) = R_0\hat{R} + Z_0\hat{Z}$, so solving that equation means solving two coupled equations, one for $dR/d\varphi$ and one for $dZ/d\varphi$.  

Equation (\ref{delta-dot}) for $\vec{\delta}$ is obtained from the equation for neighboring magnetic field lines, which solve the exact equation $d(\vec{x}+\vec{\delta})/d\varphi=\vec{h}(\vec{x}+\vec{\delta})$, by taking the limit as $\big|\vec{\delta}\big|\rightarrow0$.  Equation (\ref{delta-dot}) should be solved for two different initial conditions.  The first solve is for $\vec{\delta}_R = \delta_{RR}\hat{R} + \delta_{RZ}\hat{Z}$ with the initial condition  $\delta_{RR}=1$ and $\delta_{RZ}=0$.  The second solve is for $\vec{\delta}_Z = \delta_{ZR}\hat{R} + \delta_{ZZ}\hat{Z}$ with the initial condition  $\delta_{ZR}=0$ and $\delta_{ZZ}=1$.  Since Equation (\ref{delta-dot}) for the evolution of the separation $\vec{\delta}$ is linear, the initial separation can be taken to be unity without loss of generality.   

The Jacobian matrix for the Lagrangian position vector $\vec{x}(R_0,Z_0,\varphi)$ is defined by
\begin{eqnarray}
\frac{\partial\vec{x}}{\partial\vec{x}_0} &\equiv& \left(\begin{array}{cc}\frac{\partial R}{\partial R_0} & \frac{\partial R}{\partial Z_0} \\\frac{\partial Z}{\partial R_0} & \frac{\partial Z}{\partial Z_0}\end{array}\right), \mbox{   so  } \\
&=&\left(\begin{array}{cc} \delta_{RR} & \delta_{RZ}  \\\ \delta_{ZR}  & \delta_{ZZ} \end{array}\right).
\end{eqnarray}
The determinant of the Jacobian matrix, $\delta_{RR}\delta_{ZZ}- \delta_{ZR}\delta_{RZ}$ is the Jacobian $\mathcal{J}=RB_\varphi(R_0,\varphi_0,Z_0)/R_0B_\varphi(R,\varphi,Z)$ if there were no numerical errors.  This follows from the magnetic field being divergence free.  The magnetic flux contained in an infinitesimal tube of area $\Delta R_0 \Delta Z_0$ is $\vec{B}\cdot\vec{\nabla}\varphi \mathcal{J} \Delta R_0 \Delta Z_0$.

The Frobenius norm of a matrix is the square root of the sum of the squares of the matrix elements and is also equal to the square root of the sum of the squares of the singular values of a Singular Value Decomposition (SVD) of the matrix.  The Jacobian of a matrix is the product of its singular values, and a $2\times2$ matrix has two singular values $\Lambda_u$ and $\Lambda_s$; by definition $\Lambda_u\geq\Lambda_s$.

Consequently, the Frobenius norm of the Jacobian matrix, $\| \partial\vec{x}/\partial \vec{x}_0\|$, gives the large singular value, $\Lambda_u$, of a Singular Value Decomposition (SVD) of the matrix $\partial\vec{x}/\partial\vec{x}_0$:  
\begin{eqnarray}
\left\| \frac{\partial\vec{x}}{\partial \vec{x}_0} \right\| &\equiv& \sqrt{ \delta_{RR}^2+\delta_{RZ}^2+\delta_{ZR}^2 +\delta_{ZZ}^2}\label{Frob-exp}\\ 
 & =& \sqrt{\Lambda_u^2+1/\Lambda_s^2} 
\end{eqnarray}
since $\Lambda_u\Lambda_s=RB_\varphi(R_0,\varphi_0,Z_0)/R_0B_\varphi(R,\varphi,Z)$, the Jacobian.

When the flow is chaotic, neighboring streamlines separate exponentially, and $\Lambda_u$ becomes exponentially large, which means $\Lambda_u$ is essentially equal to the Frobenius norm of the Jacobian matrix, $\| \partial\vec{x}/\partial \vec{x}_0\|$.  A full SVD analysis gives additional information, the directions in both $R_0,Z_0$ space and in $R,Z$ space in which trajectories exponentiate apart and exponentiate together.

The natural logarithm of the Frobenius norm of the Jacobian matrix can be used to define the magnitude of the exponentiation,
\begin{equation}
\sigma(R_0,\varphi_0,Z_0,\varphi)\equiv \ln\left(\sqrt{ \delta_{RR}^2+\delta_{RZ}^2+\delta_{ZR}^2 +\delta_{ZZ}^2}\right). \label{sigma-def}
\end{equation}
The Frobenius norm  involves a sum of positive numbers and is less numerically demanding than calculating the SVD or the Jacobian, which is the difference between two numbers, each of order the Frobenius norm squared.


\section{Description of divertor magnetics}

For a tokamak or a stellarator, the plasma confinement region is defined by an outermost confining magnetic surface, which encloses a toroidal magnetic flux $\Psi_o$.   The definition of the outermost confining magnetic surface is that no magnetic field lines inside the region bounded by that surface strike the walls but at least some magnetic field line trajectories that come arbitrarily close to that surface do.

On any surface defined by a constant $\varphi$, the toroidal flux enclosed by the chamber walls $\Psi_w(\varphi)$ is greater than the toroidal flux enclosed by the outermost confining surface $\Psi_0$.  Otherwise, the main plasma strikes the walls; this is called a limiter rather that a divertor configuration. For reasons of economy, $\Psi_w$ is only be moderately greater than $\Psi_0$.

The magnetic field associated with an acceptable divertor must have a number of features.  First, consider the constraint that the divertor chambers can cover only a small fraction of the area of the walls.
For this, the relevant tubes of magnetic flux must exit the enclosing wall at a point where $\vec{B}\cdot\hat{n}<0$, where $\hat{n}$ is the unit normal to the wall, and enter the wall where $\vec{B}\cdot\hat{n}>0$ and pass extremely close to the outermost confining surface.  They must pass so close to the  plasma, or to field lines that do, that they can be filled with exiting plasma by transport.  These transport processes are intrinsically weak compared to the speed plasma can flow along the tubes that intercept the walls, Section \ref{Sec:plasma properties}. 

Two conditions must be satisfied for the divertor chambers to occupy a small fraction of the wall area:  (i) The tubes that carry plasma to the divertor must have an area that is a small fraction $f_d$ of the area of the wall.  (ii) Since the divertor chambers have fixed locations, the interception locations of these tubes with the walls must be resilient  to changes in the plasma.  Remarkably both conditions are found to be satisfied when the interceptions with the walls of field lines started just outside the plasma are plotted for existing stellarator designs \cite{Strumbeger:divertor1992,HSX divertor:2018}.  Aaron Bader et al found \cite{HSX divertor:2018}:  ``\emph{only a small degree of three-dimensional shaping is necessary to produce a resilient divertor, implying that any highly shaped optimized stellarator will possess the resilient divertor property.}"

This resilience is surprising when viewed from the sensitivity of chaotic trajectories to small changes in the Hamiltonian \cite{Chaos-targeting:1990,Targeting:Meiss1995}, which has had practical applications in defining efficient trajectories for space probes.  On the other hand, optimized stellarator designs tend to have sharp edges at particular locations, which can act in a way related to the X-points, actually the X-lines, of a tokamak divertor.

Second, an acceptable divertor must be consistent with a large radiative power loss between the main plasma and the divertor.  Stellarators are not affected by the Greenwald limit \cite{Greenwald:2002} on the plasma density, which constrains tokamaks. Consequently, it is easier to operate with a divertor that radiates most of the power in stellarators than in tokamaks.

For a large radiative power loss, two conditions must be satisfied: (i) The flux tubes that form the outflow channels must be sufficiently long $\ell_d$ from where they are loaded near the outermost confining surface to the divertor chamber to allow a large change in the plasma temperature along the field lines through radiative cooling.  The second condition (ii) depends on whether the radiative cooling is in a radiative mantle or  the divertor is detached.  

A radiative mantle must cover the whole plasma surface to prevent neutrals from entering the plasma, which can charge exchange with high-energy particles in a plasma allowing high-energy neutrals to strike the walls and sputter wall material. The radiative mantle should also block the entry into the main plasma of any sputtered material or other impurities.  Although the flux tubes that carry plasma to the divertors can cover only a small fraction of the wall area, $f_d<<1$, they can cover the whole plasma surface by making many toroidal transits while only a short distance outside the outermost confining surface.

For a detached divertor, the geometry of the tubes and the walls must be consistent with a large neutral compression ratio.  The compression ratio is the density of the neutrals in the region just outside the divertor chamber divided by the average neutral density  in the region between the plasma and the chamber walls.  A large compression ratio has three benefits: (i) A high neutral density at the pumps allows the flux of particles to be high even though the particles are cold and move slowly.  (ii)  High density regions radiate more.  (iii) A low neutral density surrounding the plasma reduces wall damage due to sputtering.   A high compression ratio requires the the existing flux tubes be well columnated with the neutral back-flow limited by material baffles  or magnetic baffles such as islands filled with plasma. 


\section{Types of divertors}

Erika Strumberger introduced the two basic types of stellarator divertors in calculations for W7-X.  The first \cite{Strumbeger:divertor1992} was in 1992 for a non-resonant divertor, which does not depend on the presence of an island chain just outside of the outermost confining magnetic surface.  The second \cite{Strumberger:divertor1996} was in 1996 for a divertor that depends on the existence of an island chain that has internal flux surfaces, which enclose a toroidal flux $\Psi_I$ that is comparable to the flux in the region between the wall and the plasma $\Psi_w-\Psi_0$.  This second design was the one used for W7-X, and its properties are described in \cite{Feng:2021,W7-X:10-2021}.  A special part of the wall, called target plates, is designed to intercept both the magnetic field lines that pass close to the outermost confining surface and some of the field lines that are confined by the island chain.

As would be expected from the Hamiltonian mechanics literature 
\cite{Stochasticy:1984,Fractal-Bnd:1988}, the outermost confining surface is not smooth; it defines a fractal boundary.  Magnetic field lines just outside the outermost surface have an exponential increase in separation with distance along the lines as long as they remain close to the outermost surface.  Once they are sufficiently far from the outermost surface, the rate of exponentiation $d\sigma_\ell/d\varphi$ becomes small unless the lines pass close to an island-like structure.  An important question is the breadth of the region of rapid exponentiation, the chaotic region, compared to the distance over which plasma processes can load magnetic flux tubes with exiting particles.    A broad chaotic region implies the magnetic field lines make many toroidal turns $M_d$ between where plasma is loaded on the flux tubes and where they hit the wall; their length is $\ell_d\approx M_d 2\pi R_0$, where $R_0$ is the major radius.   The large number of toroidal turns about the plasma makes it plausible that there can be the shielding required for a radiative mantle.   The long lines also allow a large drop in temperature between the hot plasma edge and the cool  region near the divertor.

The tokamak axisymmetric divertor seems different with a sharp separtatrix, on which the rotational transform $\iota=0$, that is smooth except at one or a small number of sharp corners called X-points.  Nevertheless, any breaking of axisymmetry breaks the separatrix into a band of chaotic magnetic field lines.  When that band is wide compared to the distance over which plasma processes can load magnetic flux tubes with exiting particles, the tokamak divertor takes on mathematical properties associated with stellarator divertors.  Even when the chaotic band is narrower, tokamak divertors resemble island divertors.  The divertors of tokamaks are often perturbed with non-axisymmetric magnetic fields in order to eliminate heat pulses due to Edge Localized Modes (ELM's), which can rapidly degrade divertor components.   As computationally predicted and demonstrated on the KSTAR tokmak \cite{Park:2018}, these perturbations can be made consistent with axisymmetric confinement in the plasma core by doing a stellarator-like optimization for quasi-axisymmetry.  A question of great practical importance to the tokamak program is can carefully chosen non-axisymmetric perturbations produce better control over the diverted plasma.  Indeed as discussed in  Section \ref{Sec:plasma properties}, it appears non-axisymmetric divertors allow control over the width of the exhaust channels.


\section{Magnetic field properties}

Although the outermost magnetic surface is not a smooth curve, the magnetic field line Hamiltonian $\psi_p(\psi_t,\theta,\varphi)$ is a smooth function.  Adjacent points on an irrational magnetic surface in arbitrarily chosen $\varphi$ plane have a mathematically subtle relationship \cite{Boozer:2022a,Boozer:2022b}.  When a field line is followed it will come arbitrarily close to every other point on the magnetic surface in a constant $\varphi$ plane.  Nevertheless, no matter how long the line is followed, it will never return to its starting point, for if it did after $M$ toroidal and $N$ poloidal transits of the torus, the surface would be a rational surface with a transform $\iota=N/M$.   

Let $\vec{x}(\varphi)$ be the trajectory of a field line that returns close to its $\varphi=\varphi_0$ starting point after $M_0$ toroidal transits.  Then, $\delta(\varphi) =|\vec{x}(\varphi) - \vec{x}(\varphi  - 2\pi M_0)|$ is small at $\varphi=\varphi_0+2\pi M_0$ and gives the separation between these two points on the magnetic surface for all $\varphi$ from plus to minus infinity.  Where the field-line separation $\delta$ is small, the neighboring magnetic surfaces have a large separation and vice versa  \cite{Boozer:2022a,Boozer:2022b}.  As the outermost surface is approached from within the region it encloses, the field-line separation $\delta(\varphi)$ appears to become erratic as does the separation between neighboring magnetic surfaces.  It is where surfaces approach each other that it is easiest for non-ideal effects such as resistivity to produce reconnection and where plasma effects can most easily move particles from one magnetic surface to another \cite{Boozer:2022a,Boozer:2022b,Boozer:2023}.  These complexities in field-line behavior remain even in a curl-free stellarator model in which the magnetic field in the region enclosed by the coils has the form $\vec{B}=\vec{\nabla}\phi$ with $\nabla^2\phi=0$ and $\phi$ a well-behaved analytic function.

As is obvious in the curl-free stellarator model, the location of the wall can be chosen to be within the region in which magnetic surfaces exist and thereby eliminate any chaotic region.  The opposite situation, a broad region of chaotic field lines, is a result of design choices, but having a chaotic region between the region of magnetic surfaces and the chamber wall seems to arise naturally from other choices made in stellarator optimization.

What is obscure is the available degree of practical control over divertor properties while remaining consistent with other optimization targets.  Extreme sensitivity, as might be expected in a chaotic system, could make practical control difficult.  But, extreme sensitivity does not appear to be the case, and some control over the divertor properties would be valuable. 

Given a stellarator design, the magnetic field line trajectories that intercept the wall can be followed within the volume enclosed by the wall until they have a second interception.   The required number of toroidal transits for a given line $M_\ell$ can be arbitrarily large, so actual studies place an upper limit on the number of transits that are followed.  For each point on the wall with  $\vec{B}\cdot\hat{n}\neq0$, one can find $M_\ell$ and $\sigma_\ell$ as well a the trajectory $\vec{x}_\ell(\varphi)$ for the magnetic field line that passes through that point.  


The strike points of a divertor are defined by lines launched from the wall that come sufficiently close to the outermost plasma surface to be loaded with particles by transport processes or sufficiently close to another field line that can.   An important parameter is, $f_d$, the fraction of the wall area that is taken up by the strike points of the divertor.

Small magnetic islands and related structures within the  region of chaotic magnetic field lines between the outermost bounding magnetic surface and the walls define apertures through which field lines must pass in going between the wall and the region outside the outermost bounding magnetic surface that can be filled with particles by transport processes.  When the region outside this outermost toroidal region that is directly filled by transport is a cantorus with openings called turnstiles \cite{Stochasticy:1984}, the divertor strike points are essentially  the image of the turnstiles on the walls.  This is an implication of magnetic of magnetic flux conservation, but there is a subtlety since tubes of magnetic flux become exponentially contorted when  $\sigma_\ell>>1$.  That is, the map of the turnstile on the wall can come arbitrarily close to points that occupy a larger area than just the area of the map, which means arbitrarily small transport processes cause the bigger area to be filled---just as in fast magnetic reconnection \cite{Boozer:2023}.   


\section{Plasma properties \label{Sec:plasma properties}}

Plasma information is required to determine what conditions must be satisfied for the chaotic region just outside the outermost surface to form an effective shielding and radiative mantle.  First, the number of toroidal circuits that field lines make while in the chaotic mantle, $M_d$, must be sufficiently large that (a) the diffusive loading of the field line reaches a depth (i) to define flux tubes of a size $\delta_d$ consistent with the size of the divertor structures. (ii)  The length of the lines in the chaotic region, $M_d 2\pi R_0$ with $R_0$ the average major radius, must be sufficiently long compared to the electron mean free path to provide adequate thermal shielding so radiative losses of energy can balance those of conductive transport.   (iii) The flux tubes carrying plasma exhaust must encircle the plasma a sufficient number of times to a provide shielding cover for the plasma of adequate thickness $\delta_s$, which turns out to be the most stringent condition on $M_d$.  The chaotic layer must be sufficiently thick to be consistent with these requirements. 

The plasma flows equally well from the region near the plasma edge to the walls on field lines that intercept the walls with $\vec{B}\cdot\hat{n}<0$ as $\vec{B}\cdot\hat{n}>0$. This is equivalent to a plasma flow with $v_{||}<0$ and $v_{||}>0$ along the lines that lead to the wall structures.  In a chaotic region, lines with these counter-streaming flows can pass close to each other, which allows even weak transport processes to remove momentum from the flow.  This process may be important for obtaining detachment \cite{Feng:2021,W7-X:10-2021} but is ignored here for simplicity as it is in a recent analysis of heat loading in the W7-X divertor \cite{Feng:2022}.


\subsection{Dependence of the diffusive width on $M_d$ }

A critical plasma parameter is the plasma cross-field diffusion coefficient.   In the EMC3-Eirene code \cite{EMC3-Eirene}, the diffusion coefficient is prescribed and is typically of order a meter squared per second.  Here we use the gyro-Bohm coefficient, $D_{gb}$, which as shown in the appendix of \cite{CO2-Stell} gives the standard scaling of confinement in tokamaks and stellarators and can be written as $D_{gb}=162 T^{3/2}/aB^2$ with the temperature in units of 10~keV, the plasma radius $a$ in units of meters, the magnetic field in Tesla, and time in seconds.  With W7-X parameters of $a=0.5~$m and $B=2.5~$T,  gyro-Bohm diffusion gives a meter squared per second at 720~eV.

The characteristic speed with which plasma flows down magnetic field lines that strike the walls is the sound speed,  $C_s=0.71\times10^6 \sqrt{T}$ in meters per second with the temperature in units of 10~keV, so the characteristic outflow time is $M_d 2\pi R_0/C_s$.  The width of exiting flux tubes that are filled with plasma by diffusion is then  
\begin{eqnarray}
\delta_d &\approx&\sqrt{D_{gb} \frac{M_d 2\pi R_0}{C_s} }\\
&\approx& 3.8\sqrt{M_d \frac{R_0}{a} \frac{T}{B^2}}~\mbox{cm}. \label{delta_d}
\end{eqnarray}
Typical parameters for a power plant might be $R_0/a=7$, $T=1~$keV at the plasma edge, and $B=5~$T, which would give $\delta_d \approx 6.3\sqrt{M_d}~$mm.


\subsection{Dependence of thermal conduction on $M_d$}

If the length of the flux tubes that carry plasma to the walls were less than an electron mean free path, heat would be carried at the electron thermal speed, which would be inconsistent with radiative cooling balancing energy transport along the magnetic field lines.  For longer flux tubes, the rate cooling by conduction along the magnetic field is the electron collision frequency, $\nu_e \approx 4.5\times 10^3n/T^{3/2}$, times the square of the ratio of the mean free path divided by the length of the lines.  The mean free path of electrons is $\lambda_e\approx10^4T^2/n$, where $n$ is the number density of electrons in units of $10^{20}~$m$^{-3}$.  The minimum number of toroidal transits of the magnetic field lines required to make $M_d2\pi R_0>\lambda_e$ is 
\begin{eqnarray}
M_c &\equiv& \frac{\lambda_e}{2\pi R_0}\\
&=& 1.6\times10^3 \frac{T^2}{nR_0}.  \label{M_c} 
\end{eqnarray}
Using $T=1~$keV, $n=10^{19}~$m$^{-3}$, and $R_0=15~$m, which are typical for a power plant, $M_c\approx10$.


\subsection{Dependence of the plasma shielding on $M_d$}

The minimum number of toroidal transits to be consistent with  shielding is 
\begin{equation}
M_s=\frac{2\pi \kappa_pa}{\delta_d} \label{M_s}.
\end{equation}
The factor $\kappa_p$ is the increase in the average poloidal circumference due to shaping.  For example, if the plasma cross section were a rectangle with sides $w$ and $h$ and $\pi a^2\equiv wh$, the area, then $\kappa_p = (w+h)/\sqrt{\pi wh}$.  When $\delta_s$, the thickness required for shielding, is larger than  the width of the plasma-filled flux tubes $\delta_d$, $M_s$ would be increased by a factor $\delta_s/\delta_d$ to obtain multiple layers of coverage.  
 
 Equations (\ref{M_s}) and (\ref{delta_d}) together imply 
 \begin{eqnarray}
 M_d \gtrsim \left(\frac{2\pi \kappa_p a}{3.8 \sqrt{\frac{R_0}{a} \frac{T}{B^2}}}\right)^{2/3}.
 \end{eqnarray}
 Using the values given after Equation (\ref{delta_d}) and $\kappa_pa=3$~m, one obtains $M_d\gtrsim208$.   A few times larger number of toroidal transits may be required to obtain adequate coverage because of overlapping of flux tubes.

The mantle must have a minimal thickness $\delta_s$ to shield the plasma.  The density of external hydrogen neutrals decays as 
\begin{eqnarray}
\frac{1}{n_H}\frac{dn_H}{dx} = - n_e \frac{I + \sigma_{cx}v_{cx}}{v_H},
\end{eqnarray}
where the ionization coefficient $I\approx4\times10^{-14}$~m$^3/$s and the charge exchange coefficient  
$\sigma_{cx}v_{cx}\approx5\times10^{-14}$~m$^3/$s, are given under various conditions in \cite{Nardon:NF2017}.  The velocity of the neutrals is $v_H\approx 2\times10^4$~m/s at an energy of 4~eV.   Consequently, the number density of neutral drops an e-fold when
\begin{eqnarray}
\int n_e dx& \approx& 2\times 10^{17}~\mbox{m}^{-2}\\
\delta_s &\equiv& \frac{\int n_e dx}{n} \approx \frac{2 \mbox{mm}}{n}  
\end{eqnarray}
is the shielding distance when the density is in units of $10^{20}~\mbox{m}^{-3}$.


\subsection{Edge density and temperature}

The ratio of the edge $n$ to the average plasma density in the core $n_c$ is 
\begin{eqnarray}
\frac{n}{n_c} \approx M_d\frac{\pi aR_0}{\kappa_p\delta_d C_s\tau_p},
\end{eqnarray}
where the particle confinement time is $\tau_p$.  Letting $a=2$~m, $\kappa_p=1.5$, $R_0=15$~m, $B=5$~T, and $\tau_p=10$~s, the ratio
$n_/n_c \approx 4.7\times 10^{-3}\sqrt{M_d}(1~\mbox{keV}/T).$
When $M_d$ is a few hundred, the edge density is approximately 10\% of the average density of the core plasma.

The ratio of the thermal energy density at the edge to the core is
\begin{equation}
\frac{W}{W_c} =\frac{a}{\kappa_p\delta_d} \frac{1}{\nu_e\tau_E}\frac{M_d^2}{M_c^2}.
\end{equation} 
This ratio is given by the relative volumes of the core and edge and their relative cooling rates.  The energy confinement time of the core plasma $\tau_E\approx 4~$s.  The rate of energy loss of the edge plasma due to conduction is approximately $\nu_e (M_c/M_d)^2$.  Keeping only the temperature $T$ and density $n$ unspecified and using the values in the previous paragraph for the others
\begin{eqnarray}
T&\approx& \left(4.3\times10^{-8}T_c\right)^{1/4} M_d^{3/8} \\
&\approx&2.6\times10^{-2}M_d^{3/8},
\end{eqnarray}
which is approximately 1.9~keV.

The thermal conductivity along magnetic field lines is proportional to $T^{5/2}$.  Consequently when $M_d>>M_c$, the electron temperature can reach arbitrarily low values along the lines if radiation becomes an important energy-loss mechanism.  Assuming the mean free path of the radiation is long compared to the local spatial scale of the plasma, the radiative power loss, Watts per meter cubed, due to atoms of atomic number $Z$ is given by $n_en_zL_z(T)$, where $n_e$ is the electron number density and $n_z$ is the number density of atoms of atomic number $Z$.  The emission rate coefficient $L_z(T)$ has a peak at a low temperature $\lesssim20~$eV.  For temperatures above this peak cooling tends to have an unstable balance, since lowering the temperature causes the radiative power loss to increase.  Consequently when radiative cooling dominates, the plasma temperature tends to drop to the maximum of $L_z(T)$.


\subsection{Consistency of requirements for $M_d$}

Adequate coverage of the plasma for shielding by the mantle appears to require a few hundred toroidal transits.  Equation (\ref{delta_d}) for the width of the exiting flux tubes then gives a reasonable width of the tubes, $\approx10$~cm.  This number satisfies the constraint $M_d>>M_c$ of Equation (\ref{M_c}), which is required to make heat transport along the field lines sufficiently small that radiation can cool the plasma before it reaches divertor structures.  Because radiative losses tend to be small at high temperatures, bistable thermal equilibria are possible: one with a high temperature at the divertor with small radiative losses and one with a low temperature with almost all of the power radiated.


\section{Methods of analysis}

There are two method of studying the magnetic properties of stellarator divertors.  One uses the magnetic fields of actual stellarator configurations \cite{Strumbeger:divertor1992,Strumberger:divertor1996,HSX divertor:2018}.  The other uses the representation of magnetic systems by maps, which is discussed in the 2022 paper by Punjabi and Boozer \cite{Punjabi2022} and the references therein.  

The use of maps to study general properties of Hamiltonian systems in regions of surface breakup is related to the 1984 study of MacKay et al \cite{Stochasticy:1984}, which introduced the concepts of cantori and turnstiles to the study of the loss of the last confining surface of the standard map as the map parameter is increased.  Their turnstiles consisted of two adjoining tubes carrying equal flux---one with lines coming inward and the other with  lines going outward across the cantorus, which resembles an irrational magnetic surface but with holes at the locations of the turnstiles.  

The 2022 paper by Punjabi and Boozer \cite{Punjabi2022} was based on a map that gives surfaces related to those of a stellarator divertor.  They found that the cantorus just outside the outermost magnetic surface had some unexpected properties:  ``\emph{The exiting and entering flux tubes can be adjacent as is generally expected but can also have the unexpected feature of entering or exiting at separate locations of the cantori. Not only can there be two types of turnstiles but pseudo-turnstiles can also exist. A pseudo-turnstile is formed when a cantorus has a sufficiently large, although limited, radial excursion to strike a surrounding chamber wall.}"  In unpublished work, Alkesh Punjabi has studied the behavior of divertor-like configurations using the large islands remaining when the parameter of the standard map is unity.  For this case, he found that three turnstiles arise, two of which have adjoining flux tubes, and one has well-separated flux tubes. He also found that when the non-resonant perturbation that generates the sharp edges on the outermost surface is turned off, only turnstiles with well-separated flux tubes arise.  He has also studied the aperture effect of islands in chaotic regions and the closeness with which field lines with oppositely directed flows come to each other between the plasma and the wall. 

\section{Discussion}

The magnetic field lines in the region between the outermost confining surface and the surrounding walls of a stellarator are the foundation of divertor design.  Further theoretical studies could clarify to what extent divertors can be optimized and controlled while maintaining an optimized plasma configuration.  Favorable divertor properties depend on have a chaotic region just outside the the outermost surface.  The chaotic region  greatly lengthens the field lines that go from just outside the outermost surface to the walls, which allows large temperature differences to be maintained between the main plasma and the walls by radiative losses.  These differences are needed to avoid intolerable power loading on the divertors. In addition, the chaotic region can allow a radiative mantle to form just outside the plasma, which not only spreads the radiative power relatively uniformly over the walls but also shields the core plasma from sputtering and impurity penetration.  Wall shaping and magnetic field design should try capture the neutrals formed by the recombining plasma in the vicinity of the divertor chamber.  This makes detachment possible, neutral pumping more efficient, and reduces the problem of sputtering. 

 
\section*{Acknowledgements}

Discussions with Alkesh Punjabi were important for the clarification of concepts discussed in this paper.

This work was supported by grant 601958 within the Simons Foundation collaboration ``\emph{Hidden Symmetries and Fusion Energy}" and by the U.S. Department of Energy, Office of Science, Office of Fusion Energy Sciences under Award Numbers DE-FG02-95ER54333 and DE-FG02-03ER54696.

\section*{Data availability statement}

Data sharing is not applicable to this article as no new data were created or analyzed in this study.




\cleardoublepage

\end{document}